\documentclass[a4paper,aps,prb,superscriptaddress,twocolumn]{revtex4}
\usepackage{amsmath,amsbsy,amsfonts,epsfig,bm}

\newcommand{\D}{\mathrm{d}}

\newcommand{\mvec}[1]{\bm{#1}}
\newcommand{\br}{{\mvec{r}}}

\newcommand{\mint}[1]{\int\! \D #1 \, }
\renewcommand{\Im}{\mathfrak{Im}}

\begin{document}

\title{Optical and magnetic excitations of metal-encapsulating Si cages:
A systematic study by time-dependent density functional theory}

\author{Micael J.~T.~Oliveira}
\email[Electronic address:\;]{micael@fis.uc.pt}
\affiliation{Center for Computational Physics, University of Coimbra, Rua Larga,
3004-516 Coimbra, Portugal}
 
\affiliation{European Theoretical Spectroscopy Facility (ETSF)}

\author{Paulo V.~C.~Medeiros}
\affiliation{Department of Physics, Chemistry and Biology (IFM), Link\"oping
University, 58183 Link\"oping, Sweden}

\author{Jos\'e~R.~F.~Sousa}
\affiliation{Center for Computational Physics, University of Coimbra, Rua Larga,
3004-516 Coimbra, Portugal}

\author{Fernando Nogueira}
\affiliation{Center for Computational Physics, University of Coimbra, Rua Larga,
3004-516 Coimbra, Portugal}
\affiliation{European Theoretical Spectroscopy Facility (ETSF)}

\author{Gueorgui K. Gueorguiev}
\affiliation{Department of Physics, Chemistry and Biology (IFM), Link\"oping
University, 58183 Link\"oping, Sweden}

\begin{abstract}
Systematic study of the optical and magnetic excitations of twelve
MSi$_{12}$ and four MSi$_{10}$ transition metal encapsulating Si cages
has been carried out by employing real time time-dependent density
functional theory. Criteria for the choice of transition metals (M)
are clusters’ stability, synthesizability, and diversity. It was found
that both the optical absorption and the spin-susceptibility spectra
are mainly determined by, in decreasing order of importance: 1) the
cage shape, 2) the group in the Periodic Table M belongs to, and 3)
the period of M in the Periodic Table. Cages with similar structures
and metal species that are close to each other in the Periodic Table
possess spectra sharing many similarities, e. g., the optical
absorption spectra of the MSi$_{12}$ (M~=~V, Nb, Ta, Cr, Mo, and W),
which are highly symmetric and belong to groups 4 and 5 of the
Periodic Table, all share a very distinctive peak at around 4 eV.  In
all cases, although some of the observed transitions are located at
the Si skeleton of the cages, the transition metal specie is always
significant for the optical absorption and the spin-susceptibility
spectra. Our results provide finger-print data for identification of
gas-phase MSi$_{12}$ and MSi$_{10}$ by optical absorption
spectroscopy.
\end{abstract}

\maketitle

\section{Introduction}

After mixed silicon-transition metal clusters (MSi$_n$) were obtained
and studied by Beck~\cite{beck87:4233} in the late eighties, the first
theoretical calculations indicated the stability of species containing
one transition metal atom such as ZrSi$_{20}$,~\cite{jn96:249}
MSi$_{15}$ (M = Cr, Mo, W),~\cite{hs01:33} and MSi$_{16}$ (M = Ti, Hf,
Zr).~\cite{kk01:045503} In 2001, a major breakthrough in the field of
silicon-transition metal clusters was achieved by Hiura et
al.~\cite{hmk01:1733} They successfully synthesized MSi$^{1+}_n$
species by employing gas phase reactions of single transition metal
atoms with silane (SiH$_4$). Abundance of clusters with 12 silicon
atoms (MSi$_{12}$) for the cases of M = Ta, W, Re, and Ir was measured
by mass-spectrometry. Initial computer simulations conducted by the
same research team suggested for MSi$_{12}$ the shape of a regular
hexagonal prism.~\cite{hmk01:1733,mhk02:121403} Subsequent theoretical
studies by others confirmed similar cage-like geometries for neutral
species such as CrSi$_{12}$,~\cite{krj02:016803} as well as for
CuSi$_{12}$, MoSi$_{12}$, and WSi$_{12}$.~\cite{hxl03:035426}

In a series of two works, in an attempt to systematize the behavior of
a wide variety of neutral MSi$_n$ clusters, we studied these
species~\cite{gp03:10313,gpsh06:1192} by means of first-principles
calculations within the framework of density functional theory
(DFT). A total of 336 different MSi$_n$ clusters, for 24 transition
elements grouped into the following subgroups: M~=~[Cr, Mo, W], [Mn,
  Tc, Re], [Fe, Ru, Os], [Co, Rh, Ir], [Ti, Zr, Hf], [V, Nb, Ta], [Ni,
  Pd, Pt], and [Cu, Ag, Au], were addressed. The geometries of most
MSi$_{12}$ and of some MSi$_{10}$ are not only symmetric, but - apart
from minor distortions - independent on the transition metal
involved. Importantly, the electronic properties of structurally
nearly equivalent clusters depend sensitively on M, thus providing
diversity of building blocks for synthesizing cluster-assembled
materials and for designing nano-devices. In order to investigate
cluster-assembled materials, i.e., solids in which the MSi$_{12}$
clusters play the role of building blocks, we also conducted Langevin
molecular dynamics simulations to investigate the possibility of
forming NbSi$_{12}$ and WSi$_{12}$ solid phases.~\cite{pgm02:033401}
In an important contribution to this emerging field, Uchida et
al. synthesized hydrogenated TaSi$_{10-13}$ clusters, successfully
deposited them onto a Si(111)-(7 x 7) surface,~\cite{ubmk03:l43} and
by taking scanning tunneling microscope (STM) observations
demonstrated that these units do not decompose after their deposition
on the Si surface. The preserved integrity of such cage-like molecules
deposited on a surface gives credibility to this first attempt for
MSi$_{12}$ manipulation and assembly into thin films and
low-dimensional structures. More recently, we have also predicted
nano-wires with metallic properties assembled from MSi$_{12}$ species
and found most stable those containing light transition metal atoms
such as Ni, Co, Ti, and V.~\cite{gsh08:170} The feasibility of
MSi$_{12}$-based nanowires has been successfully explored also by
others for M = Ni, Fe,~\cite{maf02:301,nge06:709} and even in the case
of M = Be.~\cite{skbk02:1243} Cluster-assembled materials are expected
to exhibit distinct properties from the pure substances, compounds,
and alloys of the chemical elements they contain.~\cite{jkr96:1} In
addition, silicon is the chemical element most worked out in current
microelectronics. Logically, the endohedral silicon-metal clusters and
the corresponding cluster-assembled materials based on them are
expected to exhibit distinct electronic but also optical properties
from those of existing Si-based materials. Therefore, the
MSi$_n$-based materials are seen as prospective compounds for
achieving integrated nano-opto-electro-mechanical
devices.~\cite{cl01:851,kumar06:1}

While structural and ground-state properties of MSi$_n$ clusters and
corresponding extended phases have already attracted significant
research
effort,~\cite{hmk01:1733,mhk02:121403,krj02:016803,hxl03:035426,gp03:10313,
  gpsh06:1192,pgm02:033401,ubmk03:l43,gsh08:170,maf02:301,nge06:709,skbk02:1243}
the excited state properties of MSi$_n$ units, let alone those of
possible extended phases made upon them, are just starting to be
investigated.  A very appropriate tool for approaching the excited
state properties of this type of clusters is the time-dependent
density functional theory (TDDFT).~\cite{rg84:997,mmngr2012} Indeed,
TDDFT has previously been used to study the optical properties of a
wide variety of clusters, including clusters containing
transition-metal species and hydrogenated silicon
clusters.~\cite{yb99:3809,hrs07:38,cmror08:144110,snf08:358,voc01:1813,tpa11:1096}
Nevertheless, the only previous TDDFT studies of MSi$_n$ clusters is
the recent work of He et al.~\cite{hwslw10:132}, in which the
second-order hyper-polarizabilities and the optical absorption spectra
in the UV-Visible region of (Sc - Zn)Si$_{12}$ were addressed using
TDDFT at the B3LYP and B3PW91 levels of theory.

In the present work, by applying real-time TDDFT methodology, we
report systematic results on the optical absorption spectra and
dynamical spin susceptibility of a range of stable MSi$_{12}$ (M~=~Ti,
V, Cr, Ni, Zr, Nb, Mo, Pd, Hf, Ta, W, Pt), and MSi$_{10}$ (M~=~Ni, Cu,
Ag, Au) species. The choice of the MSi$_{n}$ explored is motivated by
their structural stability (at DFT level), synthesizability, and our
determination to study a wide diversity of metals (M).

\section{Methodology and Computational Details}
\label{sec:method}

A great deal of information can be obtained about the electronic
structure of a given system by studying how it interacts with an
electromagnetic field. The mathematical objects that describe how the
electrons redistribute in a finite system after being perturbed by an
external electromagnetic field are called susceptibilities. A
susceptibility thus relates some observable of the system to a
perturbing field, and, therefore, it can be characterized by both the
type of field and the observable.  One well known example of such an
object is the polarizability of a finite system. In this case, the
perturbing field is an electrical field, the observable is the
electrical dipole of the system, and the polarizability is just the
ratio of the dipole to the field.  If the perturbing field is
frequency-dependent, the susceptibilities are referred to as being
{\em dynamical}. Important physical quantities can be related to the
susceptibilities. For example, the optical absorption cross-section
$\sigma(\omega)$ is trivially related to the imaginary part of the
dynamical polarizability $\alpha(\omega)$:
\begin{equation}
 \sigma(\omega) = \frac{4\pi\omega}{c} \Im\left\lbrace \left\langle \alpha(\omega) \right\rangle \right\rbrace\,,
\end{equation}
where the brackets denote orientational averaging.

One method of choice for calculating dynamical susceptibilities of a
wide range of physical systems is the time-dependent density
functional theory in its real-time formulation. Employing this method,
after applying the perturbation to the system, the time-dependent
Kohn-Sham equations are used to propagate the wavefunctions up to some
finite time, thus allowing the determination of any observable of the
system as a function of time. Although this method was initially
applied for calculating the polarizability,~\cite{yb96:4484} since the
methods used to solve the time-dependent Kohn-Sham equations are
independent of the applied perturbation and of the observable, this
approach is very simple to generalize to other types of
susceptibilities.~\cite{sbbn93:1601,tb00:4365,yb99:1271,veamfr09:4481}

In this work, we employ real-time TDDFT as implemented in the {\sc
  octopus} code~\cite{caoralmgr06:2465,aasoncmalarm12:233202} to
calculate the optical absorption spectra and the spin susceptibility
of MSi$_n$ clusters. Concerning the optical absorption spectra, a
dipolar perturbation:
\begin{equation}
 \delta v(\mathbf{r}, t) = - E_j x_j \delta(t)
\end{equation}
that acts at $t=0$ and equally excites all the frequencies of the
system was applied, and we kept track of the dipole moment of the
system as a function of time:
\begin{equation}
 \mvec{p}(t) = \mint{\br} \br \, n(\br, t) \,,
\end{equation}
where $n(\br, t)$ is the time-dependent electronic density. The
components of the polarizability tensor are then simply obtained from
the induced dipole $\delta \mathbf{p}(\omega)$:
\begin{equation}
 \alpha_{ij}^{[nn]}(\omega) = \frac{\delta p_i(\omega)}{E_j},
\end{equation}
where all the quantities were now moved to the more convenient
frequency domain, and the superscript $[nn]$ indicates that we are
looking at the density variation after a density perturbation.

Computing now the spin-susceptibility spectra, the applied
perturbation was made to be spin-dependent in the following way:
\begin{equation}
 \delta v(\mathbf{r}, t) = - E_j x_j  \delta(t)\sigma_z\,,
\end{equation}
where $\sigma_z$ is the usual Pauli matrix, and we kept track of the
spin-dipole moment:
\begin{equation}
  \mvec{s}(t) = \mint{\br} \br \, m(\br,t)\,,
\end{equation}
where $m(\br,t)$ is the time-dependent magnetization density. In this
case, the dynamical spin susceptibility tensor is obtained from the
induced spin-dipole $\delta \mathbf{s}(\omega)$:
\begin{equation}
 \alpha_{ij}^{[mm]}(\omega) = \frac{\delta s_i(\omega)}{E_j},
\end{equation}
where the superscript $[mm]$ indicates that we are looking at the
magnetization density variation after a perturbation of the
magnetization density. The dynamical spin susceptibility calculated by
following this scheme is just the spin contribution to the
magnetizability of the system~\footnote{The other contribution comes
  from the orbital angular momentum of the electrons.} which, in turn,
contains information about the spin-dependent electronic excitations
of the system.~\cite{sbbn93:1601,tb00:4365}

Before performing the TDDFT calculations, the geometries of the
MSi$_n$ clusters (in their ground state) were optimized at the DFT
level of theory by employing a plane-wave basis set for the expansion
of the Kohn-Sham orbitals. The calculations were performed by using
the VASP code.~\cite{kf96:11169} Projector Augmented-Wave (PAW)
potentials~\cite{blochl94:17953,kj99:1758} were employed and the
Generalized Gradient Approximation (GGA) was adopted. The
exchange-correlation functional chosen is the PBE (Perdew, Burke and
Ernzerhof).~\cite{pbe96:3865} A cut-off of 300 eV was used for the
kinetic energies of the plane waves included in the basis set, and a
Gaussian smearing scheme was employed to set the partial occupancies
of electronic states, with a width of 0.05 eV. The level of theory
employed in the present work has been demonstrated to be successful
for addressing the structural and electronic properties of a wide
range of nano-structured
systems.~\cite{xbjzs03:8621,xbs03:486,gp03:241401,gp01:6068,gfhsh06:374,
  fgchbsh08:191,rsmg10:16367}

The geometries of the MSi$_n$ cages were fully relaxed until the
absolute value of the largest projection component of the
Hellmann-Feynman forces acting on the atoms became smaller than
1x10$^{-2}$ eV/\AA. The convergence of the self-consistent electronic
cycles was considered to have been achieved when both the Kohn-Sham
eigenvalues and the total energies calculated in two consecutive
iterations differed by less than 10$^{-5}$ eV.

In the case of the TDDFT calculations, the core electrons were treated
using norm-conserving pseudopotentials of the Troullier-Martins
type.~\cite{tm91:1993} For some of the species (Ti, V, Cr, Nb, and
Mo), the generation of accurate pseudopotentials required the
inclusion of semi-core states in the valence space. For these cases,
the corresponding extension of the Troullier-Martins
scheme~\cite{rpm03:155111} as implemented in the Atomic
Pseudopotentials Engine (APE) code~\cite{on08:524} was used. In {\sc
  octopus}, all the relevant functions are discretized in a real-space
regular rectangular grid, and we chose the simulation box to be
composed of spheres around each atom.  Therefore, there are
essentially two parameters that control the convergence of the
spectra: the grid spacing and the radius of the spheres. We found that
a radius of 4.5 \AA\ and spacings of 0.10 \AA\ (M = Ti, V, Cr, Ni,
Cu), 0.13 \AA\ (M = Zr, Nb, Mo), and 0.14 \AA\ (M = Pd, Ag, Hf, Ta, W,
Pt, Au) were required in order to achieve a convergence of better than
0.1 eV in the spectra. Since some of the cage geometries exhibit
several spatial symmetries, it was possible to considerably reduce the
total number of calculations needed to obtain the full polarizability
tensor.~\cite{ocmr08:3392} As for the choice of the exchange and
correlation functional, it has already been shown that only small
differences are found in the excited state properties obtained within
the Local Density Approximation (LDA) and the GGA for this kind of
systems.~\cite{mcr01:3006} Therefore, all TDDFT calculations were
performed using the LDA for the exchange and correlation
potential.~\cite{pw92:13244} Nevertheless, for some selected clusters
we also performed the same calculations using the PBE functional, but
found only very minor differences in the spectra, as expected.

A “hydrogenated empty cage” Si$_n$H$_n$ can be derived from each of
the geometrically optimized MSi$_n$ clusters (n~=~12, 10). These
hydrogenated empty cages Si$_n$H$_n$ were obtained by removal of the
centrally located transition metal atom, followed by passivation with
H atoms of the dangling bonds of the remaining pure silicon cage. The
positions of the H atoms were then relaxed by employing the described
above method and convergence criteria. The positions of the silicon
atoms were kept fixed, thus preserving the skeleton of the
corresponding optimized MSi$_n$ cages.

\section{Results and Discussion}

\begin{figure}[t]
\centering
\setlength{\unitlength}{0.475\textwidth}
\begin{picture}(1.0,0.9)
\put(0.07,0.48){\includegraphics[width=0.37\unitlength]{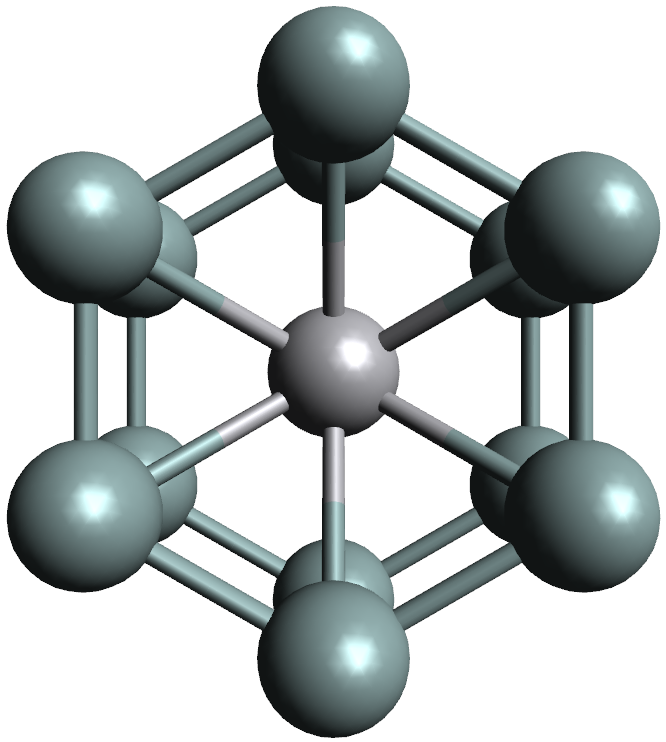}}
\put(0.54,0.50){\includegraphics[width=0.45\unitlength]{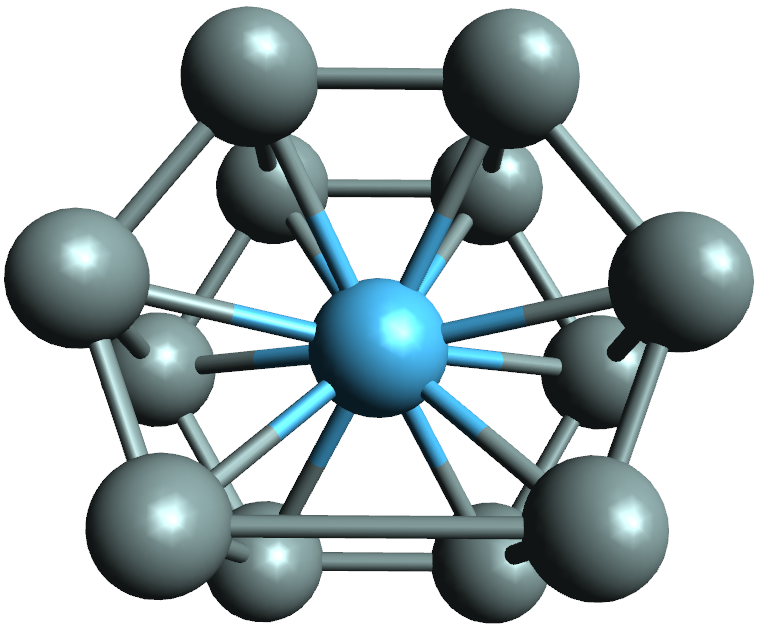}}
\put(0.05,0.02){\includegraphics[width=0.40\unitlength]{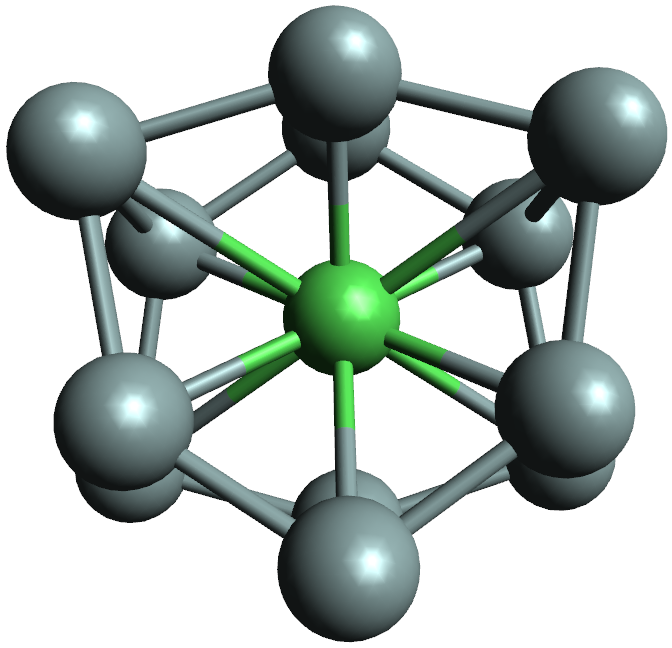}}
\put(0.55,0.00){\includegraphics[width=0.43\unitlength]{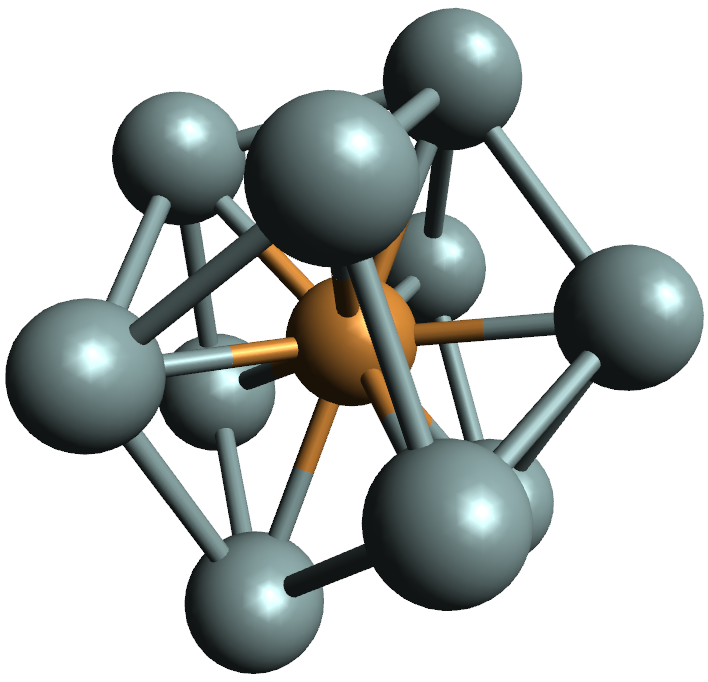}}
\end{picture}
\caption{\label{fig:MSin_geo} The four classes of geometries found for
  the selected MSi$_{n}$ clusters.  Top left: MSi$_{12}$ perfect
  $D_{6h}$ symmetry (M~=~V, Cr, Nb, Mo, Pd, Ta, W, Pt); Top right: the
  distortion of the $D_{6h}$ symmetry characteristic for M~=~Ti, Zr,
  Hf; Bottom left: the distortion of the $D_{6h}$ symmetry as for
  M~=~Ni; Bottom right: the distortion of the perfect $D_{4h}$
  symmetry MSi$_{10}$ (M~=~Ni, Cu, Ag, Au).  }
\end{figure}

The detailed study of the structural properties of these clusters has
already been done elsewhere.~\cite{gp03:10313,gpsh06:1192} As such,
here we will only briefly present our results for the ground-state
geometries of the cages insofar as these can influence the optical
properties.  Indeed, the presence of the metal atom contributes to the
optical properties of the cluster in a direct way, through its own
electronic structure and its bonds with the silicon atoms, but also in
an indirect way, as it will also influence the geometry of the
cluster. Concerning this later contribution, the overall shape of the
cluster and its symmetries are usually more important for its optical
properties than the actual values of the bond-lengths. In particular,
the spectral structure of more symmetrical clusters tends to be
simpler than the spectral structure of clusters with less symmetries,
as the number of available electronic transitions will be reduced by
the degeneracies introduced by the symmetries. On the other hand,
changes in bond-lengths that keep the overall shape typically only
introduce small shifts of the peaks of the observed spectra, and
change their relative intensities.

\begin{figure*}[ht]
  \centering
  \includegraphics[width=\textwidth]{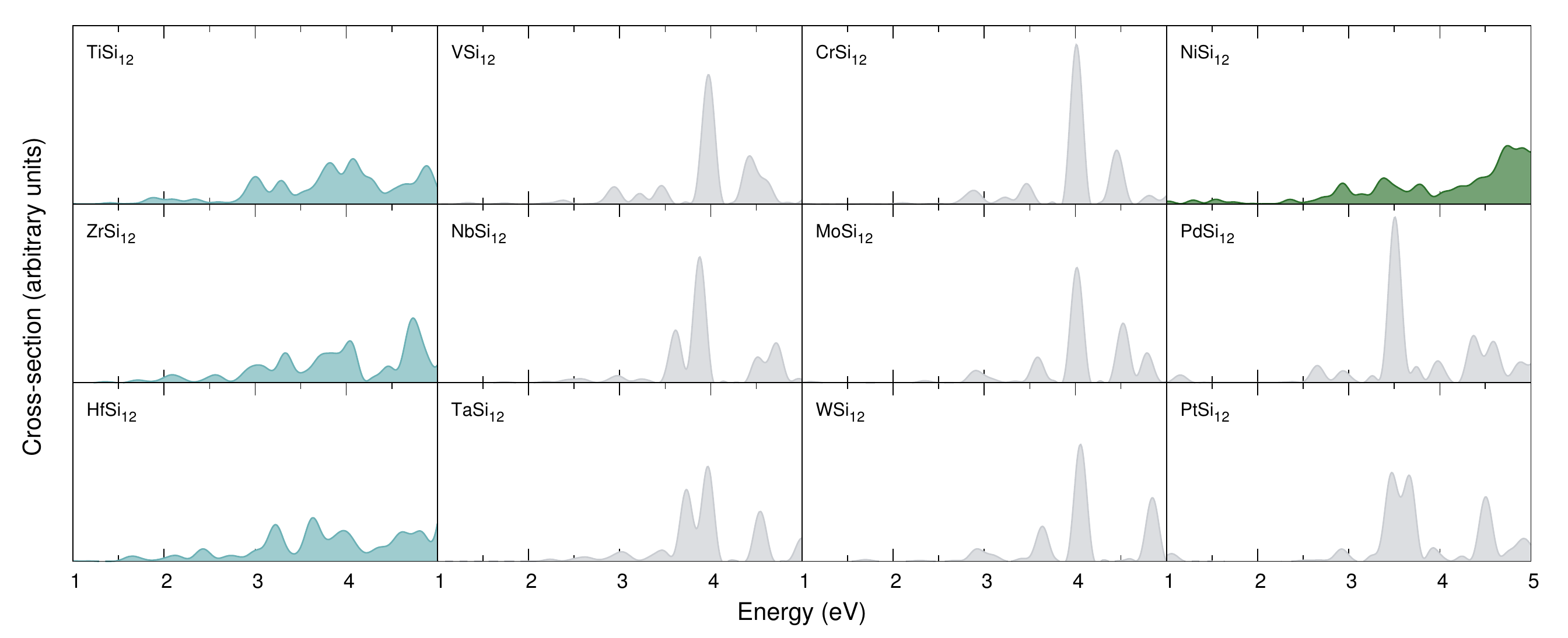}
  \includegraphics[width=\textwidth]{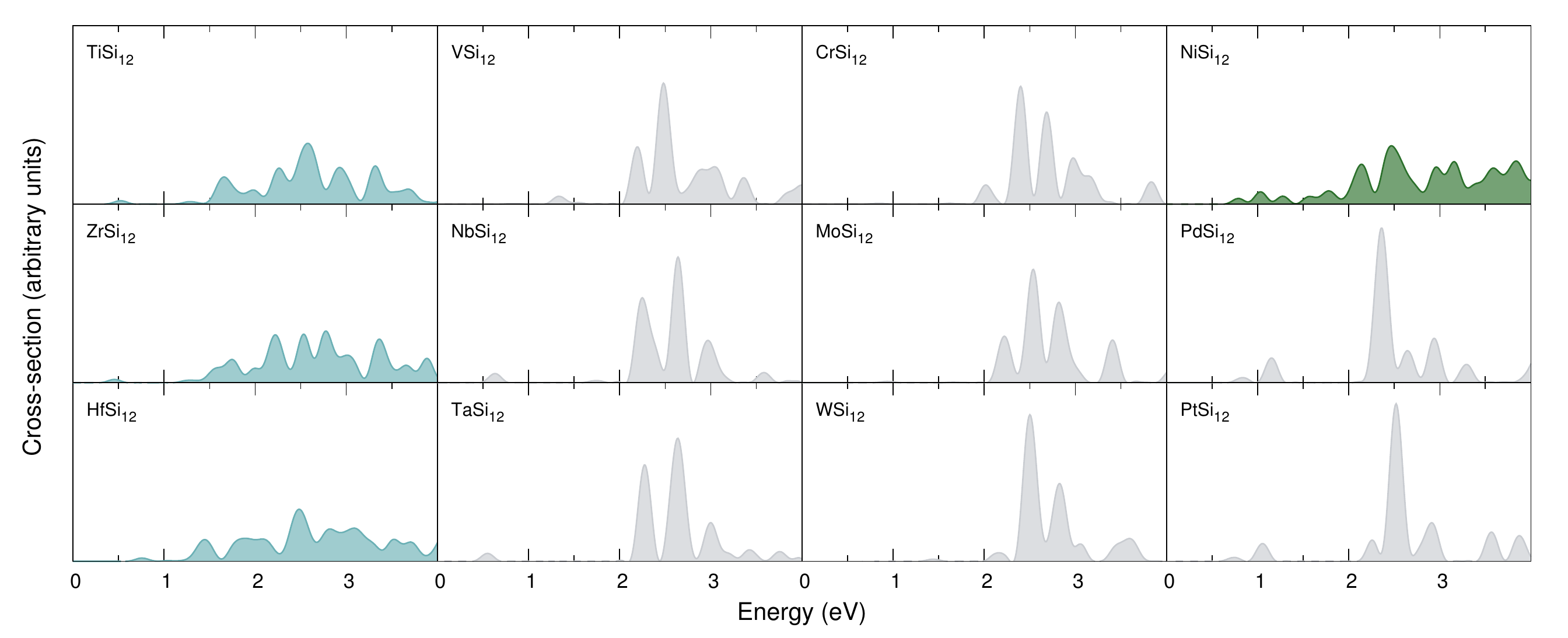}
  \caption{(color online) Optical absorption spectra (top panel) and
    spin-susceptibility spectra (lower panel) of selected MSi$_{12}$
    clusters. The color of each curve indicates the type of geometry
    of the corresponding cage: The light gray corresponds to the
    perfect hexagonal prism with $D_{6h}$ symmetry, while the blue and
    dark green correspond to the two distortions of the $D_{6h}$
    symmetry (see Fig.~\ref{fig:MSin_geo} for the equilibrium
    geometries of each of these classes of MSi$_{12}$ clusters).}
  \label{fig:optics_MSi12}
\end{figure*}

Therefore, in order to make the relationship between the metal species
and the computed excited-state properties more obvious, the obtained
geometries for the selected MSi$_{12}$ clusters were divided into
three classes. The first class corresponds to geometries with perfect
$D_{6h}$ symmetry and includes the clusters with M~=~V, Cr, Nb, Mo,
Pd, Ta, W, and Pt.  The second class of geometries includes the
clusters with M~=~Ti, Zr, and Hf, while the third class includes the
cluster with M~=~Ni. Both the second and third classes correspond to
two different distortions of the perfect $D_{6h}$ symmetry. In the
case of the selected MSi$_{10}$ clusters, there is only one class of
geometries, which corresponds to a small distortion of the perfect
endohedral bicapped tetragonal antiprism ($D_{4h}$). Representative
equilibrium geometries for the three differences classes of MSi$_{12}$
clusters as well as for the single class of MSi$_{10}$ clusters are
shown in Fig.~\ref{fig:MSin_geo}.

\subsection{MSi$_{12}$ clusters}

\begin{figure*}[ht]
  \centering
  \includegraphics[width=\textwidth]{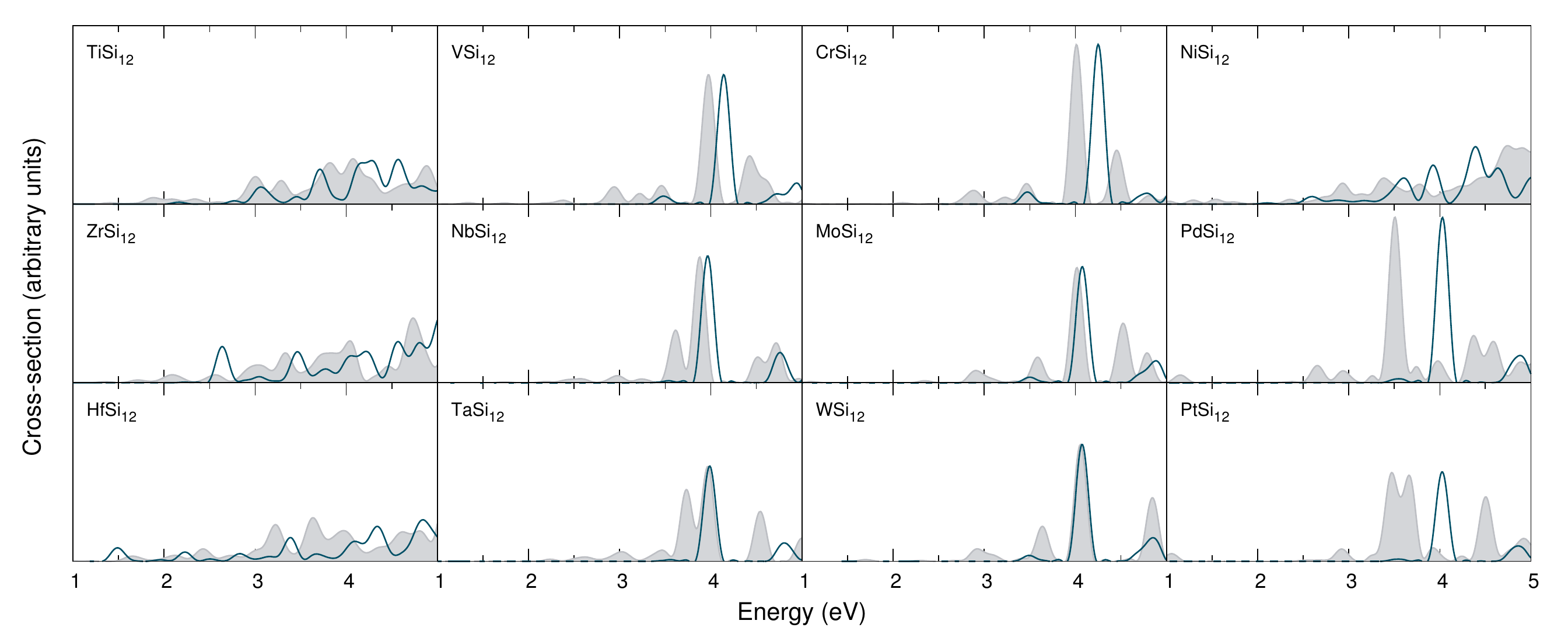}
  \caption{(color online) Optical absorption spectra of selected
    MSi$_{12}$ clusters (light gray curves) and their corresponding
    ``empty'' hydrogenated Si$_{12}$H$_{12}$ clusters (blue curves).
    The scales of the Si$_{12}$H$_{12}$ curves were adjusted to make
    the intensities of their highest peak coincide with the highest
    peak of the corresponding MSi$_{12}$ curves.}
  \label{fig:full_vs_empty_MSi12}
\end{figure*}

In Fig.~\ref{fig:optics_MSi12}, we present the calculated optical
absorption and spin-susceptibility spectra of the MSi$_{12}$ clusters
studied in this work. In these plots, the spectra are arranged with
respect to each other in a similar way as the corresponding metal
species are arranged in the Periodic Table. As for the color of the
curves, it is used to differentiate the shapes of the cages.

As seen from upper panel of Fig.~\ref{fig:optics_MSi12}, the optical
absorption spectra of cages with perfect $D_{6h}$ symmetry (light gray
curves) share significant similarities. Looking in more detail at
those spectra, we see that the ones corresponding to cages whose metal
species belong to groups 4 and 5 of the Periodic Table (VSi$_{12}$,
NbSi$_{12}$, TaSi$_{12}$, CrSi$_{12}$, MoSi$_{12}$, and WSi$_{12}$)
all share a very distinctive peak at around 4 eV. A similar peak seems
to be present in the absorption spectra of the group 10 cages
(PdSi$_{12}$ and PtSi$_{12}$), albeit shifted towards lower energy at
around 3.5 eV and, in the case of the PtSi$_{12}$, split in two peaks
close in energy. These differences can be explained by the fact that
the metal species belonging to groups 4 and 5 (V, Nb, Ta, Cr, Mo, and
W) are close to each other in the Periodic Table, while the metal
species belonging to group 10 (Pd and Pt) are further away.  A smaller
peak in the visible region, at around 3 eV, also seems to be shared by
all the spectra of these MSi$_{12}$ clusters. Other peaks are common
to several of these spectra, but not to all. We note that such similar
features are usually shared by cages for which the metal species are
close to each other in the Periodic Table. For example, the VSi$_{12}$
and CrSi$_{12}$ absorption spectra are very similar. The same applies
to the NbSi$_{12}$ and TaSi$_{12}$ cages. At this point, it is worth
to consider how the optical absorption spectra change along a given
group of the Periodic Table, as elements from the same group have
similar electronic configurations. Concerning the elements from group
4 (VSi$_{12}$, NbSi$_{12}$, and TaSi$_{12}$), we see that, with the
exception of the peak at around 4 eV, all the other peaks are slightly
shifted towards higher energies when increasing the atomic
number. There is also a peak that appears at around 3.6 eV in the
optical absorption spectra of NbSi$_{12}$ which increases in intensity
when moving to TaSi$_{12}$. As for the elements from group 5
(CrSi$_{12}$, MoSi$_{12}$, and WSi$_{12}$), the most noticeable
changes in their spectra are found in two peaks at around 3.5 eV and
4.5 eV that shift towards higher energies. In the former case, the
peak relative intensity also seems to increase when going from
CrSi$_{12}$ to WSi$_{12}$. Finally, concerning the elements from group
10 (PdSi$_{12}$ and PtSi$_{12}$), besides the aforementioned peaks at
around 3.0 and 3.5 eV, we also note the existence of similar peaks at
around 1 and 4 eV, although slightly shited towards lower energies in
the case of PtSi$_{12}$.

The same kind of analysis can also be provided for the
spin-susceptibility spectra shown in the lower panel of
Fig.~\ref{fig:optics_MSi12}, although in this case the similarities
are clearly more pronounced among spectra of cages for which the metal
species belong to the same group.  Concerning this point, we note that
the total magnetic moments of the cages with perfect $D_{6h}$ symmetry
are the same for metal elements that belong to the same group, but
different for cages for which the metal elements belong to different
groups (1 $\mu_b$, 0 $\mu_b$, and 2 $\mu_b$ for groups 4, 5, and 10 of
the Periodic Table, respectively). Looking in more detail at the
spin-susceptibility spectra of the group 4 cages (VSi$_{12}$,
NbSi$_{12}$, and TaSi$_{12}$), we observe that the structure of the
NbSi$_{12}$ and TaSi$_{12}$ spectra are almost identical, while some
differences are found for the VSi$_{12}$ spectra outside the 2-3 eV
range. These differences include the lowest energy peak, which is
shifted almost 1 eV towards higher energy in the case of the
VSi$_{12}$ spectra, and the relative intensities of the peaks found in
the 3-3.5 eV range.  Nevertheless, there are three peaks between 2 and
3 eV that feature in the spectra of all these cages. We also note that
the relative intensity of the peak at around 2.2 eV increases with the
increasing of the atomic number of the metal species. As for the group
5 cages (CrSi$_{12}$, MoSi$_{12}$, and WSi$_{12}$), three peaks are
also found between 2 and 3 eV in all cases, but there is no clear
trend regarding their relative intensities. Finally, in the case of
the group 10 cages (PdSi$_{12}$ and PtSi$_{12}$), three peaks are
again found between 2 and 3 eV for all cases with no clear trend
regarding their relative intensities. Furthermore, two peaks with
similar intensities are found between 0.8 and 1.2 eV, although
slightly shifted towards lower energies in the case of PtSi$_{12}$.

Concerning the more distorted MSi$_{12}$ clusters belonging to the
second and third classes of geometries shown in
Fig.~\ref{fig:MSin_geo}, we note that both their optical absorption
and spin-susceptibility spectra are clearly distinct from the ones of
the clusters with perfect $D_{6h}$ symmetry. In particular, the
spectra of these cages have a more complex structure, as expected.
Furthermore, we note that the change of the metal species within the
group 3 of the Periodic Table (TiSi$_{12}$, ZrSi$_{12}$, and
HfSi$_{12}$), which correspond to the clusters belonging to our second
class of geometries, does change both the optical absorption and
spin-susceptibility spectra in a noticeable way. It is possible that
the same peaks appear in the spectra of these three clusters, but the
complexity of the spectra does not allow us to ascertain it.

From the discussion above, it is clear that the cage geometry is
important in determining the excited-state properties of these
clusters. Since the spectra of cages with the same geometry share
significant spectral features, it is quite likely that some of the
calculated excitations only involve states localized at the cage
itself, i.e., at the silicon atoms. In order to verify this
hypothesis, we calculated the optical absorption spectra of the
corresponding ``empty'' hydrogenated cages, using the geometries
obtained as explained in Sec.~\ref{sec:method}. This also allows for a
better understanding of the direct contribution of the metal atom to
the excited-state properties of the clusters.

In Fig.~\ref{fig:full_vs_empty_MSi12}, a comparison between the
optical absorption spectra of the MSi$_{12}$ clusters and their
corresponding hydrogenated ``empty'' cages is provided. Looking first
at the spectra of the cages with perfect $D_{6h}$ symmetry, we note
the presence of a distinct peak located between 4.0 and 4.3 eV in all
the Si$_{12}$H$_{12}$ spectra.  These are quite close to the similar
peak found at around 4.0 eV in the case of the corresponding
MSi$_{12}$ clusters, with the notable exception of PdSi$_{12}$ and
PtSi$_{12}$, where the peak is found at around 3.5 eV. Nevertheless,
it is well noticeable that in the case of the group 4 and group 5
cages (VSi$_{12}$, NbSi$_{12}$, TaSi$_{12}$, CrSi$_{12}$, MoSi$_{12}$,
and WSi$_{12}$), the position of this peak changes a lot more in the
case of the corresponding Si$_{12}$H$_{12}$ clusters than in the case
of the MSi$_{12}$ clusters. The fact that the peak position changes in
the case of the Si$_{12}$H$_{12}$ clusters is to be expected, as the
distance between the Si atoms increases with increasing atomic number
of the metal species of the corresponding MSi$_{12}$ cluster. This
raises the question of why a similar thing does not happen in the case
of the corresponding MSi$_{12}$ clusters. A possible explanation would
be that the presence of the metal atom counteracts the change in the
spectra induced by the stretching of the Si bonds. A peak at 3.5 eV is
also shared by all the Si$_{12}$H$_{12}$ cages, although in the case
of the Nb and Ta ones, its relative intensity is quite small. We find
a peak with similar relative intensity in the spectra of VSi$_{12}$
and CrSi$_{12}$. A similar peak also appears in the spectra of
MoSi$_{12}$ and WSi$_{12}$, but at higher energies and with higher
relative intensities. From the results presented in
Fig.~\ref{fig:full_vs_empty_MSi12}, we conclude that some of the
observed transitions are, indeed, located at the silicon atoms, but
that the presence of the metal atom cannot be neglected. Once again,
the complexity of the spectra of the remaining two classes of
equilibrium geometries (TiSi$_{12}$, ZrSi$_{12}$, HfSi$_{12}$, and
NiSi$_{12}$) complicates their analysis, making it difficult to assign
specific peaks to the cage itself.

\subsection{MSi$_{10}$ clusters}

\begin{figure}[t]
  \centering
  \includegraphics[width=0.48\textwidth]{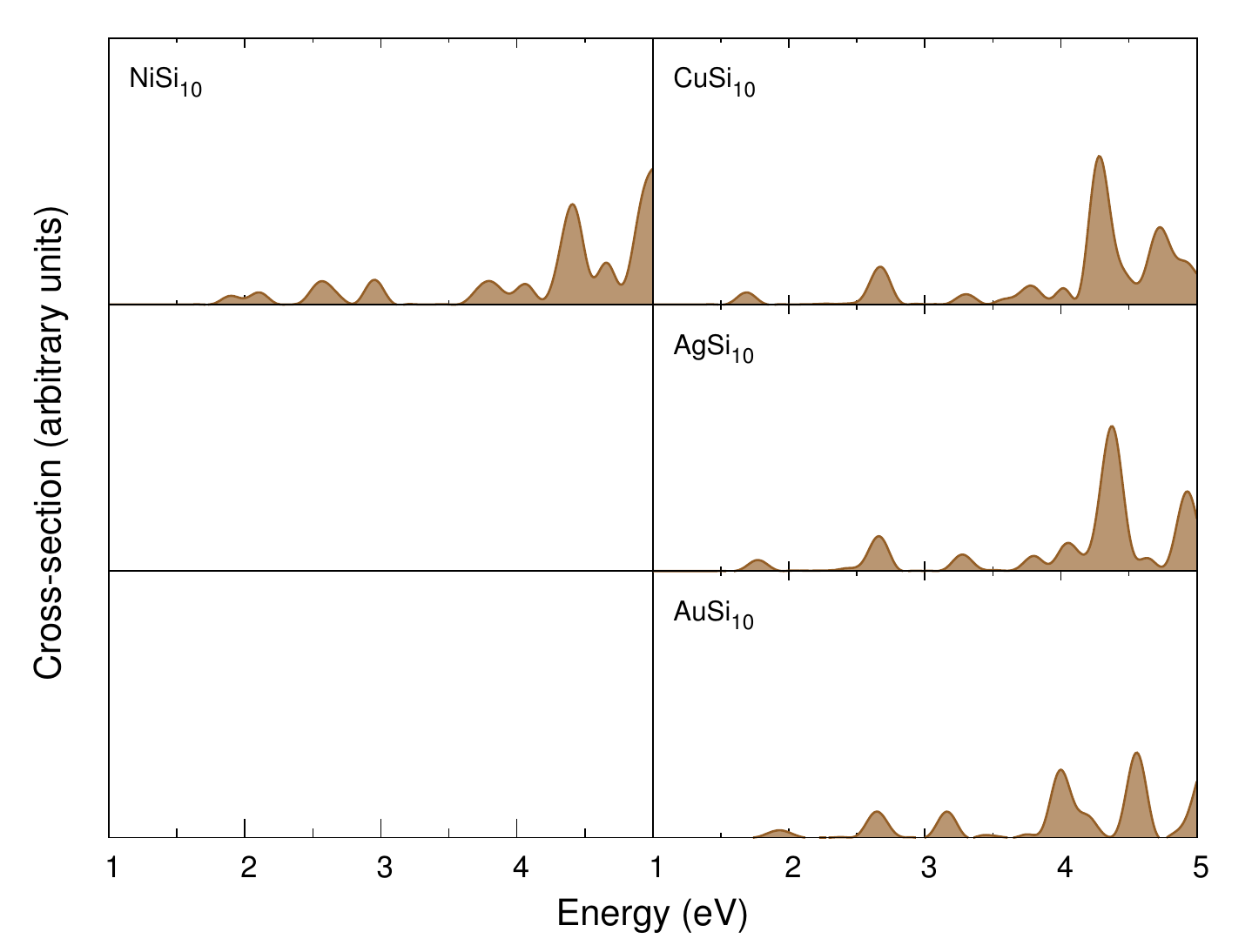}
  \includegraphics[width=0.48\textwidth]{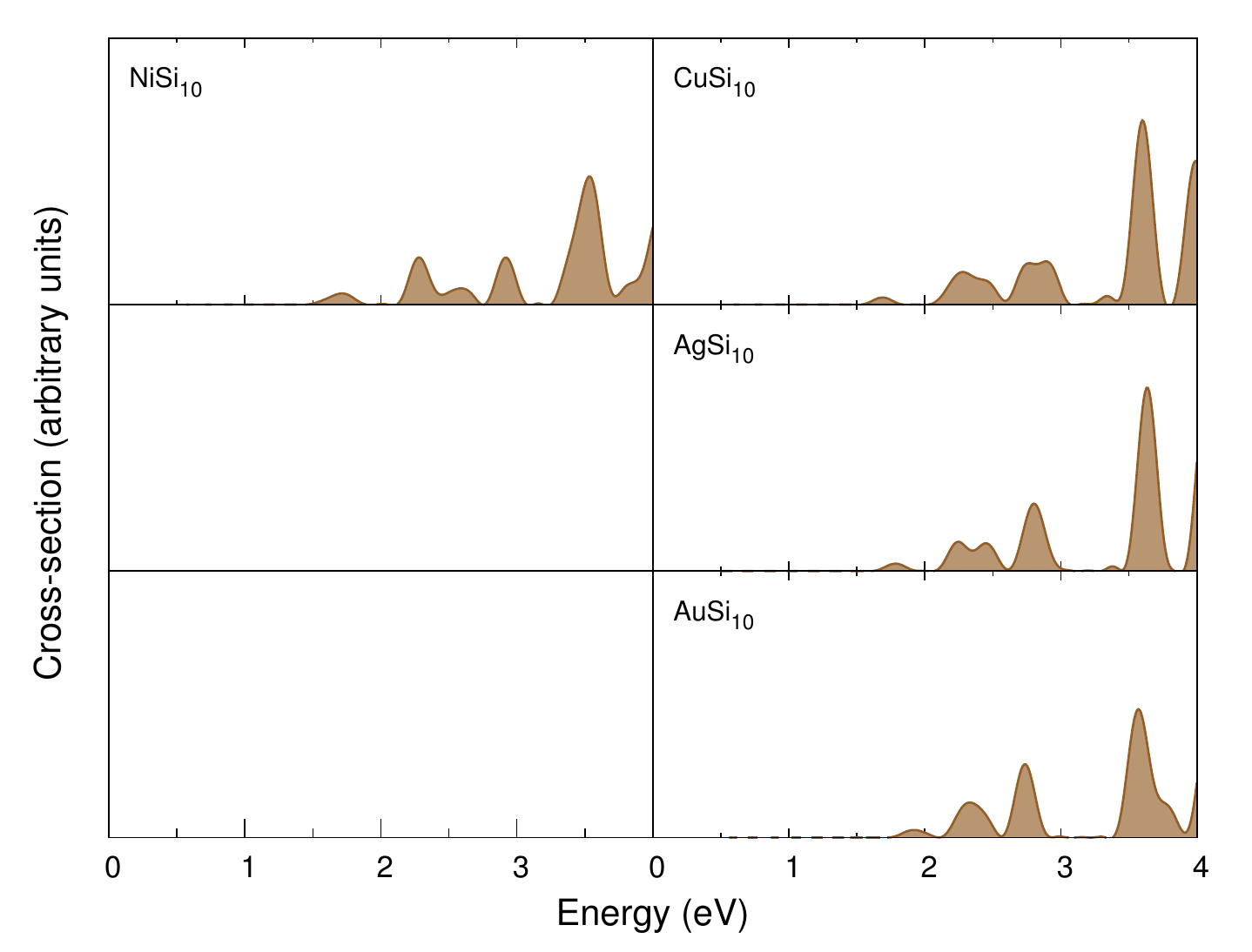}
  \caption{(color online) Optical absorption spectra (top panel) and
    spin-susceptibility spectra (lower panel) of selected MSi$_{10}$
    clusters.}
  \label{fig:optics_MSi10}
\end{figure}

In Fig.~\ref{fig:optics_MSi10}, the calculated optical absorption and
spin-susceptibility spectra of the MSi$_{10}$ clusters are shown. Like
for the MSi$_{12}$ species, the spectra are arranged with respect to
each other in a similar way as the corresponding metal species are
arranged in the Periodic Table.

From Fig.~\ref{fig:optics_MSi10}, several clear trends can be
identified in the optical absorption spectra of the group 11 cages
(CuSi$_{10}$, AgSi$_{10}$, and AuSi$_{10}$). When going from
CuSi$_{10}$ to AuSi$_{10}$, i.e., when going to heavier metal species,
we observe that the peak found between 1.5 and 2.0 eV is shifted
towards higher energies, the peak around 2.6 eV remains practically
unchanged, and the peak found between 3.0 and 3.5 eV is shifted
towards lower energies and its relative intensity increases. For the
spectral features above 3.5 eV the situation is not so clear. Despite
this, we note that the overall shapes of the CuSi$_{10}$ and
AgSi$_{10}$ spectra are quite similar, which is not the case when
comparing these spectra to the one of AuSi$_{10}$. Similarly, in the
case for the spin-susceptibility spectra, we observe that the peak
found between 1.5 and 2.0 eV is shifted towards higher energies when
going to heavier metal species.  As for the two peaks found between
2.0 and 2.5 eV in the spin-susceptibility spectra, we observe that the
peak at higher energy is shifted towards lower energies when going to
heavier metal species, eventually leading to the merging of these two
peaks in the AuSi$_{10}$ spectra.  A similar situation is also
observed for the two peaks found between 2.5 and 3.0 eV. Finally, the
peak found around 3.6 eV remains practically unchanged when going from
CuSi$_{10}$ to AgSi$_{10}$, while a splitting of the same peak seem to
occur when going from AgSi$_{10}$ to AuSi$_{10}$. Comparing the
NiSi$_{10}$ spectra with the ones of the group 11 cages, we find many
similarities. For the optical absorption spectra, we find peaks at
around 2.6, 3.7, 4.0, and 4.3 with similar relative intensities for
all the four clusters NiSi$_{10}$, CuSi$_{10}$, AgSi$_{10}$, and
AuSi$_{10}$.  As for the spin-susceptibility spectra, we find peaks
close to 1.7, 2.3, 2.8, and 3.5 eV with similar relative intensities
in all cases.  These similarities can be explained by the similar
geometry of these cages.

\begin{figure}[t]
  \centering
  \includegraphics[width=0.48\textwidth]{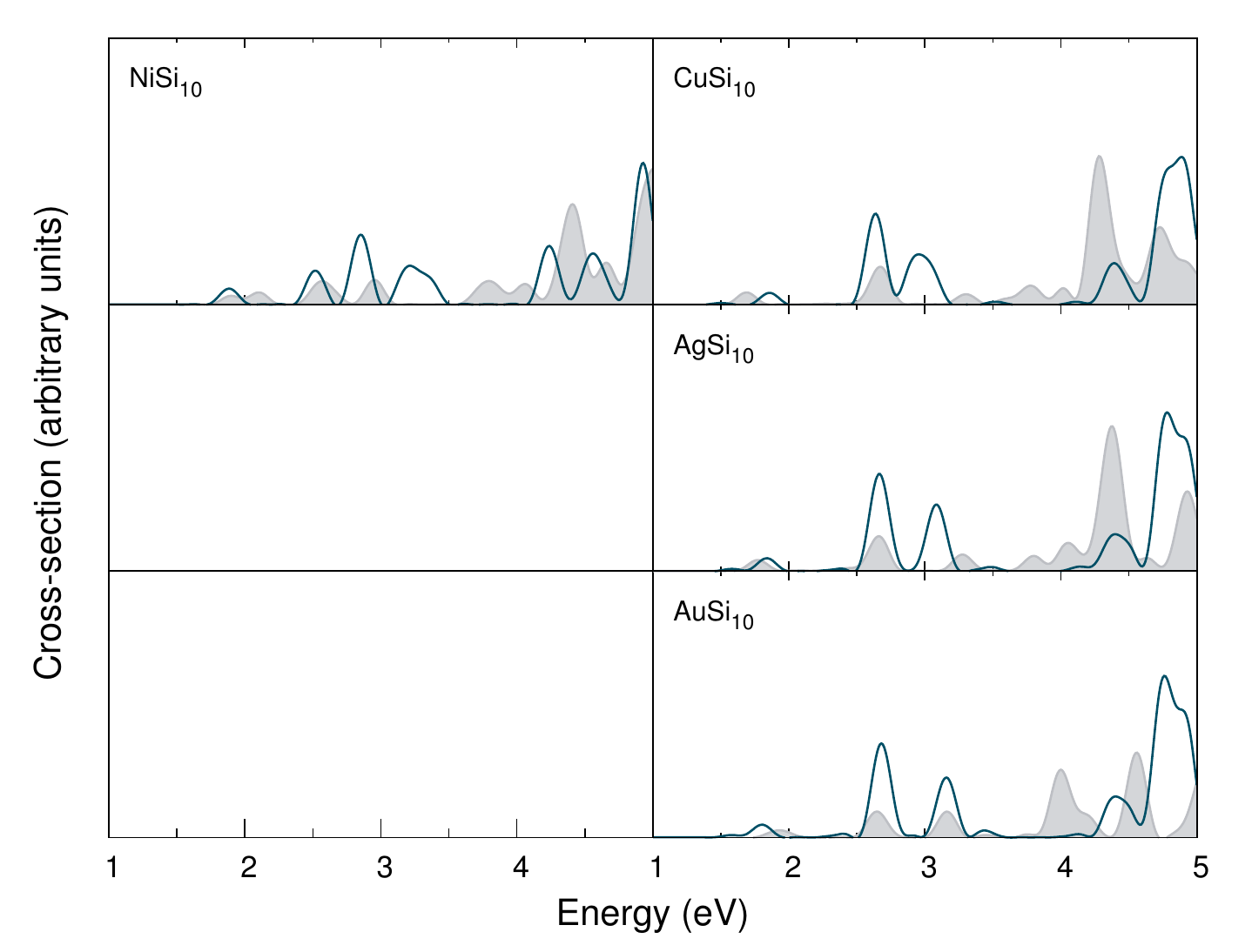}
  \caption{(color online) Optical absorption spectra of selected
    MSi$_{10}$ clusters (light gray curves) and their corresponding
    ``empty'' Si$_{10}$H$_{10}$ clusters (blue curves). The scales of
    the Si$_{10}$H$_{10}$ curves were adjusted to make the intensities
    of their highest peak coincide with the highest peak of the
    corresponding MSi$_{10}$ curves.}
  \label{fig:full_vs_empty_MSi10}
\end{figure}

As in the case of the MSi$_{12}$ clusters, we also calculated the
optical absorption spectra of the corresponding ``empty`` hydrogenated
cages for the MSi$_{10}$ clusters. The results can be seen in
Fig.~\ref{fig:full_vs_empty_MSi10}. The spectra of the
Si$_{10}$H$_{10}$ clusters are more complex than those of MSi$_{10}$,
but it is still possible to identify peaks shared with the spectra of
the MSi$_{10}$ clusters. We observe such peaks at around 1.7, 2.5 eV,
and 3.1 eV, although their relative intensities are not quite the
same. We also point out that the spectra of the Si$_{10}$H$_{10}$
cages corresponding to group 11 metal species (CuSi$_{10}$,
AgSi$_{10}$, and AuSi$_{10}$) are almost identical, while some
differences are found with respect to the spectra corresponding to the
NiSi$_{10}$ cluster. This confirms our previous conclusion drawn with
respect to the MSi$_{12}$ clusters that some of the observed
transitions are, indeed, located at the silicon atoms, but that the
presence of the metal atom cannot be neglected.

\section{Conclusions}

In this work, we have studied optical and magnetic excitations of
selected endohedral MSi$_{12}$ and MSi$_{10}$ clusters. The geometries
of the clusters were determined using density functional theory, while
a real-time formulation of time-dependent density functional theory
was used to calculate their optical absorption and spin-susceptibility
spectra.  The obtained results were then analyzed by taking into
account the position of the metal species in the Periodic Table.  From
the calculated spectra, we found that these are mainly determined by,
in decreasing order of importance: 1) the cage shape, 2) the group in
the Periodic Table the metal specie belongs to, and 3) the period of
the metal species in the Periodic Table.  This means that cages with
similar structures and metal species that are close to each other in
the Periodic Table will have spectra that share many
similarities. This implies that the fine tuning of the excited-state
properties for such cages might be better achieved by replacing the
metal atom by one belonging to the same group or, to a lesser extent,
to the same period, provided that this substitution does not alter the
shape of the cage.  In this respect, the MSi$_{10}$ clusters (M~=~Ni,
Cu, Ag, Au) and the MSi$_{12}$ clusters with perfect D$_{6h}$ symmetry
(M~=~V, Cr, Zr, Nb, Mo, Pd, Ta, W, Pt) studied in this work seem to be
particularly suited as tunable building blocks for self-assembled
nano-opto-electro-mechanical devices.  Finally, we point out that our
results indicate that optical absorption spectroscopy could be a
useful tool for the structural identification of MSi$_{10}$ and
MSi$_{12}$ clusters produced in experiments.

\begin{acknowledgments}

We thankfully acknowledge the computer resources provided by the
Laboratory for Advanced Computation of the University of Coimbra. MJTO
thankfully acknowledges financial support from the Portuguese FCT
(contract \#SFRH/BPD/44608/2008). PVCM and GKG gratefully acknowledge
support by the Swedish Foundation for International Cooperation in
Research and Higher Education (STINT) - Project YR2009-7017, and also
by the Swedish Research Council (VR). GKG acknowledges support by the
Link\"öping Linnaeus Initiative on Novel Functionalized Materials (VR)
as well as by Carl Tryggers Foundation for scientific research.

\end{acknowledgments}

\bibliography{cages}

\end{document}